\documentclass [12pt] {article}
\usepackage{a4}
 
\usepackage{amsfonts}
\usepackage{amsmath} 
\usepackage{graphics}
\usepackage{epsfig}

\author{T.~Leonhardt and W.~R\"uhl}
\title{General graviton exchange graphs for four point functions in
the AdS/CFT correspondence}

\begin{document}

\newcommand {\eps}{\varepsilon}
\newcommand {\ti}{\tilde}
\newcommand {\D}{\Delta}
\newcommand {\G}{\Gamma}
\newcommand {\de}{\delta}
\newcommand {\al}{\alpha}
\newcommand {\la}{\lambda}
\thispagestyle{empty}

\noindent hep-th/0210195   \hfill  October 2002 \\                   
 
\noindent
\vskip3.3cm
\begin{center}
 
{\Large\bf General graviton exchange graph for four point functions
in the AdS/CFT correspondence}
\bigskip\bigskip\bigskip
 
{\large Thorsten Leonhardt and Werner R\"uhl}
\medskip
 
{\small\it Department of Physics\\
     Erwin Schr\"odinger Stra\ss e \\
     University of Kaiserslautern, Postfach 3049}\\
{\small\it 67653 Kaiserslautern, Germany}\\
\medskip
{\small\tt tleon,ruehl@physik.uni-kl.de}
\end{center}
 
\bigskip 
\begin{center}
{\sc Abstract}
\end{center}
\noindent
In this note we explicitly compute the graviton exchange graph for
scalar fields with arbitrary conformal dimension $\D$ in arbitrary
spacetime dimension $d$. This results in an analytical function in
$\D$ as well as in $d$.

\newpage


\section{Introduction}
In the last years the Maldacena conjecture
\cite{Maldacena:1997re,Gubser:1998bc,Witten:1998qj} gave some deep
insight into the structure of strongly coupled gauge theories. The
possibility of exploiting the strong coupling regime lies in the
description of this theory in this regime by a dual theory, which is
weakly coupled in some limit and is therefore tractable by perturbation
theory.  The currently most investigated version of Maldacena's
conjecture is the correspondence between type II B string theory on
AdS$_5 \times S^5$ or its tractable limit II B supergravity in the
AdS$_5 \times S^5$ background, which appears for $N \rightarrow
\infty$ and $\alpha'
\rightarrow 0$, on the one hand and the $N \rightarrow \infty$ and
$\lambda \rightarrow \infty$ limit of four-dimensional $\mathcal{N}=4$
supersymmetric Yang-Mills theory with gauge group $SU(N)$ on the
other.

Besides this model derived from superstring theory there are two other
models derived by Maldacena from considerations of M2/M5 branes in
eleven dimensional M-theory. One obtains a correspondence between
eleven dimensional supergravity in the background of AdS$_{4/7} \times
S^{7/4}$ and the three and six dimensional conformal field theories on
the boundaries of AdS$_4$ and AdS$_7$, respectively.

However, as noted in \cite{Witten:1998qj} and proved in the context of
algebraic quantum field theory in \cite{Rehren:1999jn}, one can also
view AdS/CFT as a general feature of quantum field theory,
independently of string or M-theory. Appealing to this argument one
can consider field theory models on a $d+1-$dimensional AdS space and
use the AdS/CFT correspondence to define $d-$dimensional conformal
field theories. 

In conformal field theory, especially in the two-dimensional case, one
is used to handling models without any reference to a Lagrangian only
by considering correlation functions.
 There are many results on the topic of correlation functions in
 AdS/CFT in the literature
 \cite{Freedman:1998tz,Liu:1998th,D'Hoker:1998mz,Hoffmann:2000mx},
 which can be taken independently of string or M-theory. Thus we have
 a recipe for the definition of conformal field theories, at least
 perturbatively.

The aim of this note is the calculation of the ``graviton exchange
graph'', which describes the interaction of four scalar fields of
conformal dimension $\D$ in $d$ spacetime dimensions via the exchange
of a graviton in the bulk of the AdS space. This graph was already
explicitly computed in the beginning of AdS/CFT in
\cite{D'Hoker:1999pj} for the case of $d=\D=4$, because this is the
interesting case for $\mathcal{N}=4$ supersymmetric Yang-Mills
theory. Since the method of \cite{D'Hoker:1999pj} and the refining in
\cite{D'Hoker:1999ni} seems to work only for integer conformal and
spacetime dimensions, we have altered the way of computation to
include the cases of general $\D$ and $d$.

Our main motivation for the calculation of this general graviton
exchange graph is the investigation of structural issues of the
AdS/CFT correspondence as a tool for defining conformal field
theories. Nevertheless it has applications in the above mentioned
superconformal field theory derived from M2 branes, because AdS/CFT predicts
among others half integers for the dimensions $\D_{CPO}$ of the chiral primary
operators \cite{Aharony:1999ti}.


\section{Calculation of the graviton exchange graph}
\subsection{The problem and the strategy}
 
We want to calculate the following amplitude:

\begin{figure}[htb]
\begin{centering}
\resizebox{6cm}{!}{
\includegraphics{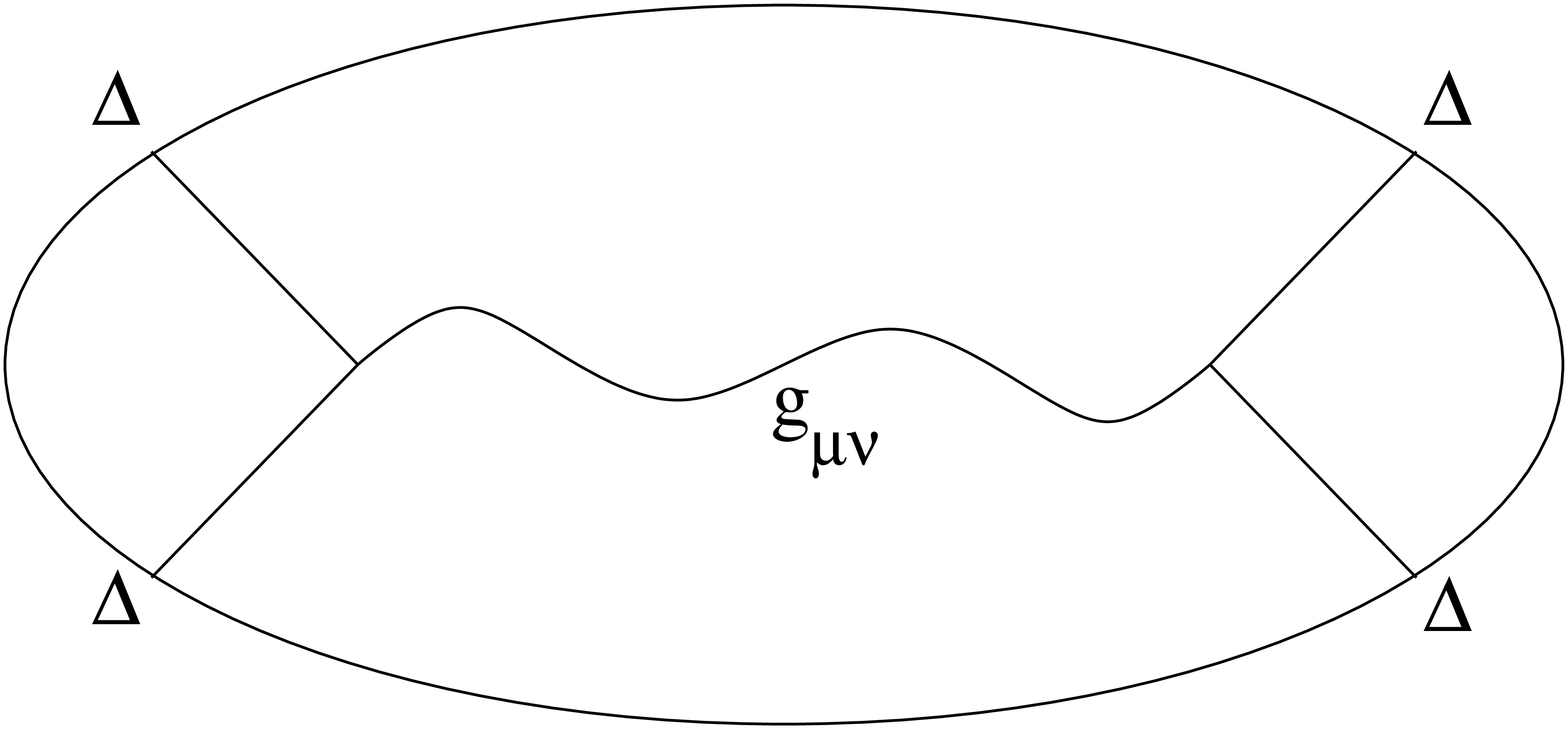}}
\caption{Witten graph of the graviton exchange amplitude}
\end{centering}
\end{figure}

Starting point is the following result taken from \cite{D'Hoker:1999ni}:
\begin{align}
G(\vec x_1,\vec x_2,\vec x_3,\vec x_4) = \tilde c \int \frac{
d^{d+1}w}{w_0^{d+1}} A^{\mu \nu}(w,\vec x_1, \vec x_3) T_{\mu
\nu}(w,\vec x_2, \vec x_4),
\end{align}
where we use the notation of the above reference and $ \tilde c$
denotes the normalization of the graviton propagator. There the factor
$A^{\mu \nu}$ is calculated and the result is 
\begin{align}
A^{\mu \nu}(w,\vec x_1, \vec x_3) = |\vec x_{13}|^{-2 \Delta}
\frac{\partial {w'}^{\lambda}}{\partial w^{\mu}} \frac{\partial
{w'}^{\rho}}{\partial w^{\nu}} I_{\lambda \rho} (w'- \vec x_{13}'),
\end{align}
with 
\begin{align}
I_{\mu \nu}=g_{\mu \nu} \frac{1}{1-d} \phi(t) + \frac{\delta_{0 \mu}
\delta_{0 \nu}}{w_0^2} \phi(t) + \textrm{gauge part},
\end{align}
where the primes denote the inverted and by $\vec x_1$ translated
coordinates and
\begin{align}
t & = \frac{w_0^2}{w^2} = \frac{w_0^2}{w_0^2 + \vec w^2},\\
\phi(t) & = -\frac{\Delta}{2} \Bigl(\frac{t}{t-1}\bigr)^{\frac{d}{2}-1}
\int_1^t dt' \, {t'}^{\Delta-\frac{d}{2}} (t'-1)^{\frac{d}{2}-2} \\ \nonumber
& = \frac{\Delta}{2} \bigl(\frac{t}{1-t}\bigr)^{\frac{d}{2}-1} \biggl\{
\frac{t^{\Delta-\frac{d}{2}+1}}{\Delta-\frac{d}{2}+1} F \Bigl[{2-\frac{d}{2},\Delta-\frac{d}{2}+1
\atop \Delta-\frac{d}{2}+2}; t \Bigr] \\ \nonumber
& \qquad\qquad\qquad\qquad\qquad -\frac{1}{\Delta-\frac{d}{2}+1} F \Bigl[
{2-\frac{d}{2}, \Delta -\frac{d}{2}+1 \atop \Delta-\frac{d}{2}+2}; 1 \Bigl] \biggr\}.
\end{align}
As usual the gauge part can be neglected, because it gives no
contribution to the integral. Thus we obtain after insertion 
\begin{align} \label{amp1}
G(\vec x_1,\vec x_2,\vec x_3,\vec x_4)= \tilde c \, \bigl( |\vec x_{13}|
|\vec x_{12}| |\vec x_{14}| \bigr)^{-2 \Delta} \int \frac{
d^{d+1}w'}{{w'}_0^{d+1}} I^{\mu \nu}(w' - \vec x'_{13}) T_{\mu
\nu}(w',\vec x'_{12}, \vec x'_{14}).
\end{align}
Now we write the last integral as $I_1+I_2$, where $I_1$ denotes the
term containing $g^{\mu \nu} T_{\mu \nu}$ and $I_2$ the term with
$T_{00}$. Explicit expressions for $g^{\mu \nu} T_{\mu \nu}$ and
$T_{00}$ are given in \cite{D'Hoker:1999pj}:
\begin{align} \label{Tgausgravex}
g^{\mu \nu} T_{\mu \nu}(w, & \vec x, \vec y) 
 = \Bigl( \frac{1-d}{2} \,\Box_w -2 m^2 \Bigr) \Bigl(
\frac{w_0}{w_0^2+|\vec w -\vec x|^2} \Bigr)^{\Delta} \Bigl(
\frac{w_0}{w_0^2+|\vec w -\vec y|^2} \Bigr)^{\Delta} \nonumber \\
& = \Bigl( (1-d)(\Delta^2+m^2) - 2m^2 \Bigr)  \Bigl(
\frac{w_0}{w_0^2+|\vec w -\vec x|^2} \Bigr)^{\Delta} \Bigl(
\frac{w_0}{w_0^2+|\vec w -\vec y|^2} \Bigr)^{\Delta} \nonumber \\
& \qquad -2(1-d) \Delta^2 |\vec x -\vec y |^2 \Bigl(
\frac{w_0}{w_0^2+|\vec w -\vec x|^2} \Bigr)^{\Delta+1} \Bigl(
\frac{w_0}{w_0^2+|\vec w -\vec y|^2} \Bigr)^{\Delta+1}
\end{align}
where we used the defining property of the bulk to boundary propagator
with $m^2=\Delta(\Delta-d)$
\begin{align}
\Box_w \Bigl( \frac{w_0}{w_0^2+|\vec w -\vec x|^2} \Bigr)^{\Delta} =
m^2 \Bigl( \frac{w_0}{w_0^2+|\vec w -\vec x|^2} \Bigr)^{\Delta}.
\end{align}
From \cite{D'Hoker:1999pj} we further learn that
\begin{align} \label{T00ausgravex}
T_{00}(w, \vec x,& \vec y) = \Delta^2 \Biggl\{
\frac{1-m^2/\Delta^2}{w_0^2} \Bigl(
\frac{w_0}{w_0^2+|\vec w -\vec x|^2} \Bigr)^{\Delta} \Bigl(
\frac{w_0}{w_0^2+|\vec w -\vec y|^2} \Bigr)^{\Delta} \nonumber \\
& -\frac{4}{w_0} \biggl[ \Bigl(
\frac{w_0}{w_0^2+|\vec w -\vec x|^2} \Bigr)^{\Delta+1} \Bigl(
\frac{w_0}{w_0^2+|\vec w -\vec y|^2} \Bigr)^{\Delta} \nonumber \\
& \qquad \qquad \qquad \quad + \Bigl(
\frac{w_0}{w_0^2+|\vec w -\vec x|^2} \Bigr)^{\Delta} \Bigl(
\frac{w_0}{w_0^2+|\vec w -\vec y|^2} \Bigr)^{\Delta+1} \biggr] \nonumber \\  
&  + \bigl( 8 + \frac{2}{w_0^2}|\vec x-\vec y|^2 \bigr)\Bigl(
\frac{w_0}{w_0^2+|\vec w -\vec x|^2} \Bigr)^{\Delta+1} \Bigl(\frac{w_0}{w_0^2+|\vec w -\vec
y|^2} \Bigr)^{\Delta+1} \Biggr\}. 
\end{align}
Now we insert (\ref{Tgausgravex}) and (\ref{T00ausgravex}) into
(\ref{amp1}), expand the hypergeometric funtion into a series and obtain
\begin{align} \label{fullgraph}
G(\vec x_1,&\vec x_2,\vec x_3,\vec x_4) =  \tilde c \, \bigl( |\vec x_{13}|
|\vec x_{12}| |\vec x_{14}| \bigr)^{-2 \Delta}
\frac{\Delta^3}{2\Delta-d+2} \nonumber \\
& \Biggl\{ \bigl(2+\frac{2}{d-1} \frac{m^2}{\Delta^2}
\bigr) \biggl[ \mathbf{M}_{0,0} -\frac{\Gamma(\Delta -\frac{d}{2}+2)
\Gamma(\frac{d}{2}-1)}{\Gamma(\Delta)} \, \overline{\mathcal{M}}_{0,0} \biggr]
\nonumber \\
& \;\; -4 \biggl[(\mathbf{M}_{1,0} + \mathbf{M}_{0,1})
-\frac{\Gamma(\Delta -\frac{d}{2}+2)
\Gamma(\frac{d}{2}-1)}{\Gamma(\Delta)} (\overline{\mathcal{M}}_{1,0} +
\overline{\mathcal{M} }_{0,1}) \biggr] \nonumber \\  
& \;\; +8 \biggl[ \mathbf{M}_{1,1} -\frac{\Gamma(\Delta -\frac{d}{2}+2)
\Gamma(\frac{d}{2}-1)}{\Gamma(\Delta)}\, \overline{\mathcal{M}}_{1,1} \biggr] 
\Biggr\}, 
\end{align}
where we have defined
\begin{align}
\mathbf{M}_{s,\ti s} := \sum_{k\ge0}\frac{(2-\frac{d}{2})_k
(\D-\frac{d}{2}+1)_k}{k!\, (\D-\frac{d}{2}+2)_k } \mathcal{M}_{k,s,\ti
s}.
\end{align} 
together with the ``standard integrals''
\begin{align}
\mathcal{M}_{k,s,\tilde s} & = \int \frac{d^{d+1}w}{w_0^{d+1}}\, \Bigl(
\frac{w_0^2}{|\vec w -\vec x'_{13}|^2} \Bigr)^{\frac{d}{2}-1} \Bigl(
\frac{w_0^2}{w_0^2+|\vec w -\vec x'_{13}|^2} \Bigr)^{\Delta-\frac{d}{2}+1+k}
w_0^{s+\tilde s}
\nonumber \\ 
& \qquad \qquad \qquad \quad \Bigl( \frac{w_0}{w_0^2+|\vec w -\vec x'_{12}|^2} \Bigr)^{\Delta+s}\Bigl(
\frac{w_0}{w_0^2+|\vec w -\vec x'_{14}|^2} \Bigr)^{\Delta+\tilde s}
\nonumber \\
\overline{\mathcal{M}}_{s,\tilde s} & =  \int
\frac{d^{d+1}w}{w_0^{d+1}}\, \Bigl( \frac{w_0^2}{|\vec w -\vec
x'_{13}|^2} \Bigr)^{\frac{d}{2}-1} w_0^{s+\tilde s} \nonumber \\ 
& \qquad \qquad \qquad \quad \Bigl( \frac{w_0}{w_0^2+|\vec w -\vec x'_{12}|^2} \Bigr)^{\Delta+s}\Bigl(
\frac{w_0}{w_0^2+|\vec w -\vec x'_{14}|^2} \Bigr)^{\Delta+\tilde s}.
\end{align}
Now the remaining problem to be solved is the calculation of these
integrals. We will give the solution as a power series in some conformally
invariant standard variables defined below (\ref{biharm}).


\subsection{Calculation of $\overline{\mathcal{M}}$}
We start with the integral $\overline{\mathcal{M}}$. The denominators
are handled with the usual Feynman parameters and we obtain after
performing the $d+1$-dimensional $w$-integral
\begin{multline}
\overline{\mathcal{M}}_{s, \tilde s} = \frac{\pi^{\frac{d}{2}}}{2}
\frac{\Gamma(\Delta+s+ \tilde s-1) \Gamma(\Delta)}{\Gamma(\frac{d}{2}-1)
\Gamma(\Delta +s) \Gamma(\Delta + \tilde s)} \int_0^1 dt_2 dt_3 dt_4
\,\delta \Bigl(\sum_{i=2}^4 t_i -1 \Bigr) \\
 t_2^{\Delta+s-1} t_3^{\frac{d}{2}-2} t_4^{\Delta+\tilde s-1} (t_2 +
t_4)^{-(\Delta +s +\tilde s -1)} y^{-\Delta},
\end{multline}
where we have defined 
\begin{align}
y & = \sum_{i=2}^4 t_i {|\vec x'_{1i}|^2} - \Bigl( \sum_{i=2}^4 t_i \vec
x'_{1i} \Bigr)^2 \nonumber \\
& = \sum_{i<j} t_i t_j |\vec x'_{1i} - \vec x'_{1j}|^2.
\end{align}
Now we use 
\begin{align}
|\vec x'_{1i} - \vec x'_{1j}|^2 = \frac{\vec x^2_{ij}}{\vec x^2_{1i}
\vec x^2_{1j}},
\end{align}
integrate out $t_3$ (which is trivial), introduce the conformally
invariant variables
\begin{align}\label{biharm}
u := \frac{\vec x^2_{13}\vec x^2_{24}}{\vec x^2_{12}\vec x^2_{34}},
\qquad v:= \frac{\vec x^2_{14}\vec x^2_{23}}{\vec x^2_{12}\vec
x^2_{34}}
\end{align}
and change the integration variables
\begin{align}\label{newintvar}
t_2 = \sigma \rho,\qquad t_4 = \sigma (1- \rho).
\end{align} 
After that we find 
\begin{align}
y = \sigma \frac{\vec x^2_{34}}{\vec x^2_{13}\vec x^2_{14}} \bigl(
1-(1-v) \rho \bigr) \Bigl(1- \sigma \bigl(1-
\frac{\rho(1-\rho)u}{1-(1-v)\rho} \bigr) \Bigr),
\end{align}
and we can perform the resulting $\sigma$-integral to get
\begin{multline}\label{argderhyp}
\overline{\mathcal{M}}_{s, \tilde s} = \frac{\pi^{\frac{d}{2}}}{2}
\frac{\Gamma(\Delta+s+ \tilde s-1) \Gamma(\Delta)}{\Gamma(\frac{d}{2})
\Gamma(\Delta +s) \Gamma(\Delta + \tilde s)} \biggl( \frac{\vec
x^2_{34}}{\vec x^2_{13}\vec x^2_{14}} \biggr)^{-\Delta} \\
\int d\rho \,\rho^{\Delta +s-1} (1-\rho)^{\Delta+\tilde s-1} \bigl(
1-(1-v) \rho \bigr)^{-\Delta} F \Bigl[{\Delta, 1
\atop \frac{d}{2}}; 1-\frac{\rho(1-\rho)u}{1-(1-v)\rho}  \Bigr].
\end{multline}
To calculate the remaining integral, we use an analytic continuation
formula for the hypergeometric function to translate the argument of the
hypergeometric function by $1$ (eq. $9.131, 2$ of \cite{GR}). The two
resulting hypergeometric functions can then be expanded as a power series
and the $\rho$-integral can be done, again in terms of gaussian
hypergeometric functions. The result is 
\begin{align}\label{allgemMq}
\overline{\mathcal{M}}_{s, \tilde s} & = \frac{\pi^{\frac{d}{2}}}{2}
\frac{\Gamma(\Delta+s+ \tilde s-1) \Gamma(\Delta)}{
\Gamma(\Delta +s) \Gamma(\Delta + \tilde s)} \biggl( \frac{\vec
x^2_{34}}{\vec x^2_{13}\vec x^2_{14}} \biggr)^{-\Delta} \nonumber \\
& \Biggl\{ \frac{ \Gamma(\frac{d}{2}- \Delta-1)}{ \Gamma(\frac{d}{2}- \Delta)
\Gamma(\frac{d}{2}-1)} \sum_{l=0}^{\infty} \frac{B(\Delta + s+l, \Delta + \tilde s
+l) (\Delta)_l}{(\Delta-\frac{d}{2}+2)_l} \nonumber \\
& \qquad \qquad \qquad \qquad \qquad \qquad \quad u^l F \Bigl[{\Delta+l,
\Delta+s+l
\atop 2 \Delta +s+\tilde s+2 l}; 1-v  \Bigr] \nonumber \\
& \;\; + \frac{ \Gamma(\Delta+1-\frac{d}{2})}{ \Gamma(\Delta)}
\sum_{l=0}^{\infty} \frac{B(\frac{d}{2}+ s+l-1, \frac{d}{2}+ \tilde s
+l-1) (\frac{d}{2}-1)_l}{l!} \nonumber \\ 
& \qquad\qquad \qquad\qquad \qquad
u^{l+\frac{d}{2}-\Delta-1} F \Bigl[{\frac{d}{2}-1+l, \frac{d}{2}+s+l-1
\atop d +s+\tilde s+2 l-2 }; 1-v  \Bigr] \Biggr\}. \nonumber \\
\end{align}


\subsection{Calculation of $\mathbf{M}$}
Now that we have seen how to compute $\overline{\mathcal{M}}$,
the procedure for $\mathcal{M}$ is straightforward. As before, we
introduce Feynman parameters (here we need four of them), perform the
$w$-integral and integrate (trivially) over $t_1$ to obtain
\begin{align}\label{Mksst1}
\mathcal{M}_{k,s,\tilde s} & = \frac{\pi^{\frac{d}{2}}}{2}\frac{ \Gamma(2
\Delta + k+ s+ \tilde s -\frac{d}{2}) \Gamma(\Delta)}{\Gamma(\frac{d}{2}-1)
\Gamma(\Delta- \frac{d}{2} +1 +k) \Gamma(\Delta + s) \Gamma( \Delta + \tilde
s)} \nonumber \\ & 
\qquad \int dt_2 dt_3 dt_4 \; t_2^{\Delta + s-1} t_3^{\Delta +k -\frac{d}{2}}
t_4^{ \Delta + \tilde s-1} \bigl( 1- t_2 -t_3 -t_4 \bigr)^{\frac{d}{2}-2}
\nonumber \\
& \qquad \qquad \qquad \bigl(t_2+t_3+t_4 \bigr)^{-(2\Delta+k+s+\tilde
s-\frac{d}{2})} y^{-\Delta},   
\end{align}
where we used the abbreviation $y$ given above. Again we make the
substitution (\ref{newintvar}) and get for the integral over $t_3$
\begin{align}\label{t3int}
\int_0^{1-\sigma} & dt_3 \, t_3^{\Delta-\frac{d}{2}+k}
\bigl(1-\sigma-t_3\bigr)^{\frac{d}{2}-2}
\bigl(\sigma+t_3\bigr)^{-(2\Delta+k+s+\tilde s -\frac{d}{2})} \nonumber \\
& = \bigl(1-\sigma\bigr)^{\Delta+k-1} \sigma^{-(2\Delta+k+s+\tilde s
-\frac{d}{2})} B\bigl( \Delta+k+1-\frac{d}{2},\frac{d}{2}-1\bigr) \nonumber \\
& \qquad \qquad \qquad F \Bigl[{2\Delta+k+s+\tilde
s -\frac{d}{2}, \Delta+k-\frac{d}{2}+1 \atop \Delta+k} ; \frac{\sigma-1}{\sigma}  \Bigr]
\nonumber \\
& = \bigl(1-\sigma\bigr)^{\Delta+k-1} B\bigl(
\Delta+k+1-\frac{d}{2},\frac{d}{2}-1\bigr) \nonumber \\ 
& \qquad \qquad \qquad F \Bigl[{2\Delta+k+s+\tilde s -\frac{d}{2}, \frac{d}{2}-1
\atop \Delta+k} ; \frac{\sigma-1}{\sigma}  \Bigr],
\end{align}
where we used the first transformation formula of eqns. 9.131, 1 in
\cite{GR}. We plug (\ref{t3int}) back into (\ref{Mksst1}), expand the
hypergeometric function and integrate over $\sigma$ to get
\begin{align}
\mathcal{M}_{k,s,\tilde s} = & \frac{\pi^{\frac{d}{2}}}{2}
\frac{\Gamma(2\Delta+k+s+\tilde s -\frac{d}{2})\Gamma(\Delta)}{
\Gamma(\Delta+k) \Gamma(\Delta+s) \Gamma(\Delta+\tilde s)} 
\biggl(\frac{\vec x_{34}^2}{\vec x_{13}^2\, \vec x_{14}^2}
\biggr)^{-\Delta} \nonumber \\
& \quad \sum_{n\ge0} \frac{(2\Delta+k+s+\tilde s -\frac{d}{2})_n (\frac{d}{2}-1)_n}{n!\,
(\Delta +k)_n} B\bigl( \Delta+s +\tilde s, \Delta +k+n \bigr)
\nonumber \\ 
& \qquad\quad  \int d\rho\, \rho^{\Delta+s-1}
\bigl(1-\rho\bigr)^{\Delta+\tilde s -1}
\bigl(1-(1-v)\rho\bigr)^{-\Delta}  \nonumber \\ 
& \qquad \qquad \qquad F \Bigl[{\Delta, \Delta+s+\tilde s \atop
2\Delta+k+n+s +\tilde s } ; \, 1- \frac{\rho(1-\rho)u}{1-(1-v)\rho} \,
\Bigr]. 
\end{align}   
Again we apply the analytic continuation formula 9.131, 2 of
\cite{GR} to translate the argument of the last hypergeometric
function by $1$. Since we run into poles in the formulas,
\footnote{These singularities are only formal artifacts, the content
is fully analytical, of course.} we make the shift $k \mapsto k+ \eps$
and obtain after performing the last integration
\begin{align}
\mathcal{M}_{k,s,\tilde s} = & \frac{\pi^{\frac{d}{2}}}{2}
\frac{\Gamma(2\Delta+k+\eps+s+\tilde s -\frac{d}{2})}{
\Gamma(\Delta+k+\eps) \Gamma(\Delta+s) \Gamma(\Delta+\tilde s)} 
\biggl(\frac{\vec x_{34}^2}{\vec x_{13}^2\, \vec x_{14}^2}
\biggr)^{-\Delta} \nonumber \\
& \quad \sum_{n\ge0} \frac{(2\Delta+k+\eps+s+\tilde s -\frac{d}{2})_n (\frac{d}{2}-1)_n}{n!\,
(\Delta +k+\eps)_n \Gamma(\Delta+k+\eps+n+s+\tilde s)} \nonumber \\
& \qquad \Biggl\{ \Gamma(k+\eps+n) \Gamma(1-k-\eps-n) \sum_{m,l \ge 0}
\frac{u^l (1-v)^m}{l!\, m!} \nonumber \\
& \qquad \qquad \frac{\Gamma( \D + \ti s +l) \G ( \D +s +\ti
s+l)}{ \G ( 1-k-\eps -n+l)} \frac{ \G ( \D +l+m) \G ( \D+s+l+m)}{ \G (
2( \D+l) +s+\ti s+m)}\nonumber \\
& \qquad + \G(-k-\eps-n) \G(1+k+\eps+n) \sum_{m,l\ge0}
\frac{u^{k+\eps+n+l}(1-v)^m}{l!\, m!} \nonumber \\
& \qquad \qquad \frac{\G(\D+\ti s +k+\eps+n+l) \G(\D+s+\ti s
+k+\eps+n+l)}{\G(1+k+\eps+n+l)} \nonumber \\
& \qquad \qquad \frac{\G(\D+k+\eps+n+l+m)
\G(\D+s+k+\eps+n+l+m)} {\G(2(\D+k+\eps+n+l)+s+\ti s+m)} \Biggl\}.
\end{align}
The next step is the summation over $k$. We observe that up to two
factors all $k-$summands in $\mathcal{M}_{k,s,\ti s}$ solely depend on
$k+n$, therefore by introducing the new summation variable
$N:=k+n$, we shift the summation variables from $(k,n)$ to
$(k,N)$. Since the resulting $k-$sum is a $_3 F_2-$series, which can be
summed by Saalsch\"utz theorem \cite{slater}, we get
\begin{align}
\mathbf{M}_{s, \ti s} & = \frac{\pi^{\frac{d}{2}}}{2} \frac{1}{\G(\D+s)\G(\D+
\ti s)} \biggl(\frac{\vec x_{34}^2}{\vec x_{13}^2\, \vec x_{14}^2}
\biggr)^{-\Delta} \nonumber \\ 
& \; \sum_{N\ge0} \frac{\G(2\D+s+\ti s-\frac{d}{2}+N+\eps)}{\G(\D+N+\eps)
\G(\D+s+\ti s+N+\eps)} \frac{(\D)_N}{(\D-\frac{d}{2}+2)_N} \frac{\pi}{\sin
\pi(N+\eps)} \nonumber \\
& \;\; \sum_{l,m\ge0}\frac{u^l(1-v)^m}{l!\;m!} \Biggl\{
\frac{\G(\D+s+\ti s+l) \G(\D+\ti s+l)}{\G(1-N-\eps+l)} \nonumber \\
& \qquad \qquad \qquad \qquad\qquad\qquad \frac{\G(\D+l+m)
\G(\D+s+l+m)}{\G(2(\D+l)+s+\ti s+m)}\nonumber \\
& \qquad \qquad \qquad \qquad - u^{N+\eps} \frac{\G(\D+s+\ti s+N+\eps+l)
\G(\D+\ti s+N+\eps+l)}{\G(1+N+\eps+l)} \nonumber \\
& \qquad \qquad \qquad \qquad\qquad \frac{\G(\D+N+\eps+l+m)
\G(\D+s+N+\eps+l+m)} {\G(2(\D+N+\eps+l)+s+\ti s+m)} \Biggr\}.
\end{align}  
To perform the sum over $N$ we note that the first double series in
$u,1-v$  contains only one $N-$dependent term
\begin{align}
\frac{1}{\G(1-N-\eps+l)} = (-1)^N \frac{(\eps-l)_N}{\G(1+l-\eps)},
\end{align}
and in the second double series the summands only depend upon $L:=N+l$,
up to 
\begin{align}
\frac{1}{l!} = (-1)^N \frac{(-L)_N}{L!},
\end{align}
so that we obtain after replacing $L \mapsto l$
\begin{align}
\mathbf{M}_{s,\ti s} = & \frac{\pi^{\frac{d}{2}}}{2} \frac{1}{\G(\D+s)\G(\D+
\ti s)} \biggl(\frac{\vec x_{34}^2}{\vec x_{13}^2\, \vec x_{14}^2}
\biggr)^{-\Delta} \nonumber \\ 
& \; \sum_{N\ge0} \frac{\G(2\D+s+\ti s-\frac{d}{2}+N+\eps)}{\G(\D+N+\eps)
\G(\D+s+\ti s+N+\eps)} \frac{(\D)_N}{(\D-\frac{d}{2}+2)_N} \frac{\pi}{\sin
\pi\eps} \nonumber \\
& \;\; \sum_{l,m\ge0}\frac{u^l(1-v)^m}{l!\;m!} \Biggl\{
\frac{\G(\D+s+\ti s+l) \G(\D+\ti s+l)}{\G(1-\eps+l)} \,(\eps-l)_N \nonumber \\
& \qquad \qquad \qquad \qquad\qquad\qquad \frac{\G(\D+l+m)
\G(\D+s+l+m)}{\G(2(\D+l)+s+\ti s+m)}\nonumber \\
& \qquad \qquad \qquad \qquad\quad - u^{\eps} \frac{\G(\D+s+\ti s+\eps+l)
\G(\D+\ti s+\eps+l)}{\G(1+\eps+l)}\, (-l)_N \nonumber \\
& \qquad \qquad \qquad \qquad\qquad \quad\frac{\G(\D+\eps+l+m)
\G(\D+s+\eps+l+m)} {\G(2(\D+\eps+l)+s+\ti s+m)} \Biggr\}.
\end{align}  
At last we have to perform the limit $\eps \rightarrow 0$. To do this
we observe that the pole $\sin^{-1}\pi\eps$ is cancelled by the
zero of $\{\cdots\}$ at $\eps=0$, such that the result is analytical, as
promised before. We take the limit with the help of l'H\^opital's rule
and arrive finally after a short calculation at
\begin{align}\label{fettM}
\mathbf{M}_{s,\ti s} = & \frac{\pi^{\frac{d}{2}}}{2} \frac{\G(2\D+s+\ti s-\frac{d}{2})}
{\G(\D)\G(\D+s)\G(\D+\ti s)\G(\D+s+\ti s)}
\biggl(\frac{\vec x_{34}^2}{\vec x_{13}^2\, \vec x_{14}^2} \biggr)^{-\Delta} 
\nonumber \\  
& \; \sum_{l,m\ge0}\frac{u^l(1-v)^m}{l!\;m!} \frac{\G(\D+s+\ti s+l)
\G(\D+\ti s+l)}{\G(1+l)}\nonumber \\
& \qquad \qquad \qquad \quad \;\;\, \frac{\G(\D+l+m) \G(\D+s+l+m)}
{\G(2(\D+l)+s+\ti s+m)}\nonumber \\ 
&\qquad \Biggl\{ \,\!  _3F_2 \Bigl[{2\Delta+s+\ti s-\frac{d}{2}, 1, -l \atop
\Delta+s +\tilde s,\D-\frac{d}{2}+2 } ; \, 1 \,\Bigr] \biggl( -\log u +
2\psi(1+l) \nonumber \\ 
& \qquad \qquad\quad -\psi(\D+\ti s+l) -\psi(\D+s+\ti s+l) -\psi(\D+l+m)
\nonumber \\ 
& \qquad \qquad\quad -\psi(\D+s+l+m) +2\psi(2(\D+l)+s+\ti s+m)
\biggr)\nonumber \\
&\qquad \quad + \sum_{N=1}^l \frac{(2\D+s+\ti s-\frac{d}{2})_N (-l)_N} {(\D+s+\ti
s)_N (\D -\frac{d}{2}+2)_N} \sum_{p=0}^{N-1} \frac{1}{p-l} \nonumber \\   
& \qquad \quad + (-1)^l l! \frac{ (2\D+s+\ti s-\frac{d}{2})_{l+1}} {(\D+s+\ti
s)_{l+1} (\D -\frac{d}{2}+2)_{l+1}} \nonumber \\
& \qquad \qquad \qquad _3F_2 \Bigl[{2\Delta+s+\ti
s-\frac{d}{2}+l+1, 1, 1 \atop \Delta+s +\tilde s+l+1 ,\D-\frac{d}{2}+l+3 } ; \, 1
\,\Bigr] \Biggr\}.
\end{align}    

Now we only have to transform the last generalized hypergeometric
function in (\ref{fettM}) into a finite sum. This can be done in the
following way: We first apply the fundamental two-term relation
(eq. (4.3.1) in \cite{slater}) to get
\begin{multline}\label{zwischen}
 _3F_2 \Bigl[{2\Delta+s+\ti s-\frac{d}{2}+l+1, 1, 1 \atop \Delta+s +\tilde
 s+l+1 ,\D-\frac{d}{2}+l+3 } ; \, 1 \,\Bigr] \\
= \frac{\G(\D+s+\ti s+l+1) \G(\D-\frac{d}{2}+l+3) \G(l+1)}{\G(2\Delta+s+\ti
s-\frac{d}{2}+l+1) \G(l+2)^2} \\
 _3F_2 \Bigl[{\frac{d}{2}-\D, l+1, 2-s-\ti s-\D \atop l+2, l+2} ; \, 1 \,\Bigr].
\end{multline} 
One immediately recognizes that this series terminates in case of $\D
\in \mathbb{N}_0$ or $\D-\frac{d}{2} \in \mathbb{N}_0$. But we want to have a
calculable exact result for any field dimension $\D$ and any spacetime
dimension $d$, thus we now apply the fundamental three-term relation
(eq. (4.3.2.1) in \cite{slater}) to (\ref{zwischen}), where we choose 
\begin{align}
F_n(2)&= F_n(2;1,3) \nonumber \\
F_n(3)&= F_n(3;4,5)
\end{align}
in the notation of \cite{slater}. We include the prefactor of the
$_3F_2-$function in the last summand of (\ref{fettM}) and obtain 
\begin{align}
& (-1)^l l! \frac{ (2\D+s+\ti s-\frac{d}{2})_{l+1}} {(\D+s+\ti
s)_{l+1} (\D -\frac{d}{2}+2)_{l+1}} \nonumber \\
& \qquad \qquad \qquad \qquad \qquad _3F_2 \Bigl[{2\Delta+s+\ti
s-\frac{d}{2}+l+1, 1, 1 \atop \Delta+s +\tilde s+l+1 ,\D-\frac{d}{2}+l+3 } ; \, 1
\,\Bigr] \nonumber \\
& = \frac{l!\, \G(2\Delta+s+\ti s-\frac{d}{2}+l+1) \G(\D+s+\ti s) \G(\D+s+\ti
s-1)} {\G(2\Delta+s+\ti s-\frac{d}{2}) \G(\D+s+\ti s+l)^2
\G(\D-\frac{d}{2}+2)} \nonumber \\
& \qquad \qquad \G(\D-\frac{d}{2}+1)\, _3F_2 \Bigl[{2\Delta+s+\ti s-\frac{d}{2}+l+1, -l, \D-\frac{d}{2}+1 \atop
\D-\frac{d}{2}+2, \D-\frac{d}{2}+2} ; \, 1 \,\Bigr] \nonumber \\
& \quad - \frac{(l!)^2 \G(\D+s+\ti s) \G(\D+s+\ti s-1) \G(\D-\frac{d}{2}+2)
\G(\D-\frac{d}{2}+1)} {\G(2\Delta+s+\ti s-\frac{d}{2}) \G(\D+s+\ti s+l) \G(\D-\frac{d}{2}+2+l)}.
\end{align}

Finally we have got rid of all infinite series for the coefficients and we
can calculate for a given spacetime dimension $d>2$ and a given
conformal dimension $\D >\frac{d}{2}-1$ any coefficient of the power series in
$u$ and $1-v$ exactly in finitely many steps.


\section{Comparison with previous results}

In this section we compare our results with those of
\cite{Hoffmann:2000dx}, where the results of \cite{D'Hoker:1999pj} are
formulated in the same variables as we use. In the case of the dilaton
axion exchange graph $d=\D=4$ one can check that the coefficients of
the power series in $u$ and $1-v$ indeed agree, if we take care of the
normalizations of the graviton propagator and the bulk-to-boundary
propagators of the scalar fields.

In \cite{Hoffmann:2000dx} it was pointed out that the singular power
terms in the $(u,1-v)-$ex\-pan\-sion of the graviton exchange amplitude
arise from the exchange of an energy momentum tensor in the conformal
field theory interpretation. Let us see if the same applies to the
case of arbitrary $d$ and $\D$:

\begin{figure}[htb]
\begin{centering}
\resizebox{6cm}{3cm}{
\includegraphics{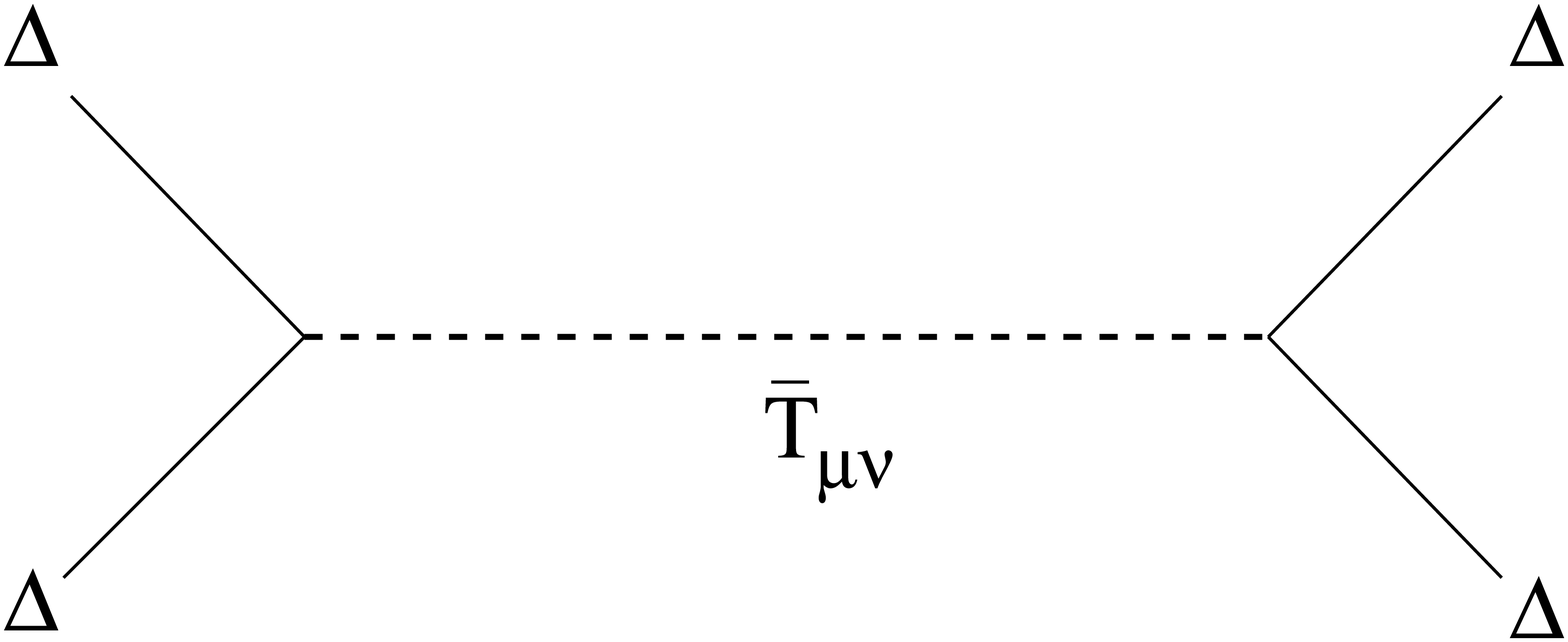}}
\caption{Conformal graph of the energy momentum exchange amplitude}
\end{centering}
\end{figure}

The exchange graph for the energy momentum tensor $\bar T_{\mu \nu}$
can be calculated with the help of the ``master formula'' of
\cite{LR}. With that formula we first give the exchange amplitude
$W_{\de}(x_1,\ldots,x_4;\D)$ for a second rank tensor of arbitrary
conformal dimension $\de$ with scalars of dimension $\D$ as exterior
legs: 
\begin{multline}
W_{\de}(x_1,\ldots,x_4;\D) = \pi^{\frac{d}{2}}
(x_{12}^2 x_{34}^2)^{-\D} C_{\D,\D} \tilde C_{\D,\D} \G(\frac{d}{2}-\de) \G(1-\frac{d}{2}+\de)
u^{\frac{\de} {2}-1-\D} \\
\sum_{n\ge0} \frac{u^n}{n!}\frac{1}{\G(\de+2n)\G(\de-\frac{d}{2}+1+n)}
\Bigl\{ \frac{1}{2} \bigl( 1-\frac{2 u}{d} \bigr) A_n(0,0)
\phi_n(0,0;1-v) \\
- \bigl[A_n(2,0)+A_n(0,2)\bigr] \phi_n(2,0;1-v) \\ + \frac{1}{2} A_n(2,2)
\bigl[\phi_n(2,2;1-v)+\phi_n(2,-2;1-v) \bigr] \Bigr\},
\end{multline}
where 
\begin{align}\label{defA}
A_n(i, j) := \Bigl(\frac{\de + i}{2}\Bigr)_n \Bigl(\frac{\de -
i}{2}\Bigr)_n \; \frac{ \G \bigl(\frac{\de + j}{2} +n \bigr) \G
\bigl(\frac{\de - j}{2} +n \bigr)} { \G \bigl( \frac{d-\de -
j}{2} \bigr) \G \bigl( \frac{d-\de + j}{2} \bigr)}
\end{align}
and
\begin{align}
\phi_n(i,j; z) := F \Bigl[ {\frac{\de-i}{2}+n,
\frac{\de-j}{2}+n \atop \de+2n }\; ;\, z \Bigr].
\end{align}
At the dimension of the energy momentum tensor $\de=d$ we recognize
that $A_n(i,0)$ has a double zero, whereas $A_n(i,2)$ has a simple
one. We renormalize the coupling constants $C_{\D,\D}, \tilde
C_{\D,\D}$ by absorbing the simple zero:
\begin{align}
C^{(r)}_{\D,\D} = \G\Bigl(\frac{d-\de - 2}{2}\Bigr)^{\frac{1}{2}} C_{\D,\D} \nonumber
\\
\tilde C^{(r)}_{\D,\D} = \G\Bigl(\frac{d-\de - 2}{2}\Bigr)^{\frac{1}{2}} \tilde C_{\D,\D}
\end{align}
Then only the terms with $A_n(i,2)$ survive the limit $\de \rightarrow
d$ and we get for the energy momentum tensor exchange amplitude
\begin{multline}\label{EI-Austausch}
W^{(r)}_{d}(x_1,\ldots,x_4;\D) = \pi^{\frac{d}{2}}
(x_{12}^2 x_{34}^2)^{-\D} C^{(r)}_{\D,\D} \tilde C^{(r)}_{\D,\D}
\frac{\G(-\frac{d}{2})}{2
\,\G(\frac{d}{2})} u^{\frac{d} {2}-1-\D}  \\ 
\sum_{n\ge0} \frac{u^n}{n!} \frac{\G(\frac{d}{2}-1+n)
\G(\frac{d}{2}+n)^2} {\G(d+2n)}  \biggl\{ -d \phi_n(0,2) \\
+\frac{(\frac{d}{2}+n) (\frac{d}{2}-1)} {(\frac{d}{2}-1+n)}
\Bigl(\phi_n(2,2)+\phi_n(2,-2) \Bigr) \biggr\}. 
\end{multline}
Note that a further renormalization is necessary, if $d$ is an even
positive integer.

The singular terms in the graviton exchange graph all stem from the
second series in $\overline{\mathcal{M}}_{s, \ti s}$ (\ref{allgemMq}). We
insert the complete second series into (\ref{fullgraph}) and obtain
after using some recursion relations for the gaussian hypergeometric
functions 
\begin{align}
G(\vec x_1,& \vec x_2, \vec x_3,\vec x_4)\Bigr\rvert_{\textrm{singular
part}} = \pi^{\frac{d}{2}} \, \tilde c \, K_\D^4(x_{12}^2 x_{34}^2)^{-\D}  
\frac{\D^2 \G(\D-\frac{d}{2}+1)^2} {\G(\D)^2 (d-1)} u^{\frac{d}
{2}-1-\D} \sum_{l\ge0} \cdots 
\end{align}
where the dots denote the same summands as in
(\ref{EI-Austausch}). Moreover, we installed the normalizations of the
bulk-to-boundary propagators $K_\D$ \cite{Freedman:1998tz}.

Thus we conclude that the second series in (\ref{allgemMq}), which
contains the singular terms, exactly coincides with the conformal
energy momentum exchange graph if we set for the coupling constant
(taking into account the above mentioned normalizing factor) 
\begin{align}
C^{(r)}_{\D,\D} \tilde C^{(r)}_{\D,\D} = \tilde c K_\D^4 \frac{\D^2
\G(\D-\frac{d}{2}+1)^2} {\G(\D)^2} \frac{2
\,\G(\frac{d}{2})}{(d-1)\G(-\frac{d}{2})}.
\end{align}

We close by mentioning that the result we have obtained is fully
analytical in the conformal dimensions $\D$ of the scalar fields as
well as in the spacetime dimension $d$. This is a general feature of
conformal field theories. E.g. the $O(N)$-sigma model has a critical
point for $2<d<4$ and the amplitudes are analytical in $d$ in this
domain. Recently a conjecture was formulated about the holographic
dual of this model in AdS \cite{Klebanov:2002ja}. Thus, the amplitudes
calculated by the AdS/CFT procedure in this model are also supposed to
show analytical behaviour in this domain.

\end{document}